\documentstyle[epsfig]{aipproc}
\def\be{\begin{equation}}
\def\ee{\end{equation}}
\def\bea{\begin{eqnarray}}
\def\eea{\end{eqnarray}}
\def\bq{\begin{eqnarray}}
\def\bq{\begin{eqnarray}}
\def\eq{\end{eqnarray}}
\begin{document}
\thispagestyle{empty}
\begin{flushright}
MCTP-01-22\\
UPRF-2000-22\\
WUE-ITP-01-09\\
\end{flushright}
\title{Exclusive Decay Amplitudes \\ from Light-Cone Sum Rules
\footnote{Talk presented by O.Y. at the Conference 
``Phenomenology 2000'', Madison, Wisconsin, April 2000}}
\author{Reinhold R\" uckl$^a$, Stefan Weinzierl$^{b}$, Oleg
Yakovlev$^{c}$}
\address{$^a$ 
Institut f\"ur Theoretische Physik und Astrophysik,\\ 
Universit\"at W\"urzburg, D-97074 W\"urzburg, Germany\\
$^b$ Dipartimento di Fisica, Universit\`a di Parma,\\
INFN Gruppo Collegato di Parma, 43100 Parma, Italy\\
$^c$
Michigan Center for Theoretical Physics,\\ 
University of Michigan, Ann Arbor, Michigan, 48109-1120}
\maketitle
\begin{abstract}
We review recent developments in QCD sum rule applications 
to semileptonic $B\to\pi$ and $D\to\pi$ transitions. 
\end{abstract} 

\section*{Introduction}
The ongoing experiments at the $B$ factories allow measurements 
of $B$ decays and $CP$ violation with greatly improved precision
(see e.g. \cite{BaBar}) which should be matched also on the 
theoretical side. The most difficult obstacle
are the long-distance QCD effects playing an important, sometimes
even dominant role in weak decays of hadrons.
One of the most powerful tools in applying QCD to  
hadron physics is the method of QCD sum rules.   
Since its invention in 1979 \cite{SVZ}, the sum rule method has become 
more and more 
advanced not only technically, but also conceptually.
A prominent example is provided by the sum rule for the semileptonic 
$B\to\pi$ transition, relevant for the determination of the CKM-matrix
element $V_{ub}$. In the following, we discuss this transition together 
with the analogous $D\to\pi$ transition in more detail. 
Measurements of the latter can serve as valuable cross checks for the 
QCD sum rule method.

The non-perturbative dynamics of the above heavy-to-light transitions 
is encoded in two form factors $f^+$ and $f^-$ parameterizing the hadronic 
transition matrix elements. Focussing on $B\to\pi l \nu_l$, one has
\bq
\langle \pi | \bar{b} \gamma_\mu q | B \rangle & = 
& 2 f^+ q_\mu + \left( f^+ + f^- \right) p_\mu,
\eq
with $q_{\mu}$ and $p_{\mu}$ being the momenta of the pion and the lepton 
pair, respectively.
For the light electron and muon channels only the form factor $f^+$ is 
relevant, whereas in the tau channel also $f^-$ plays a role.
Sometimes, it is convenient to use the scalar form factor $f^0$
given by
\bq
\label{scalarff}
f^0 & = & f^+ + \frac{p^2}{m_B^2-m_\pi^2} f^-.
\eq

In early QCD sum rule calculations of heavy-to-light form factors 
short-distance operator product expansion was applied
to suitable three-point correlation functions, 
following the original ideas of Shifman, Vainshtein and Zakharov.
However, this approach suffers from soft end-point contributions.
A significant improvement was achieved by using instead 
two-point correlation functions such as 
\bq
F_\mu(p,q)& = & i\int dx e^{ip\cdot x}
\langle \pi(q)|T\{\bar u(x)\gamma_\mu b(x),
m_b\bar b(0)i\gamma_5 d(0)\}|0 \rangle 
\eq
where the time-ordered product of operators is sandwiched between 
the vacuum and a physical pion state.
Expansion of the operator product near the light-cone $x^2=0$ then leads
to a sum over hard scattering amplitudes convoluted with pion 
distribution amplitudes of twist 2,3,4,etc. \cite{BBKR,KRW} . 
Currently these series are truncated after twist 4.  
The resulting sum rules are usually called light-cone sum rules (LCSR).

For the $B \rightarrow \pi$ transition the LCSR allows to calculate
$f^+$ and $f^0$ in the range of momentum transfers
$0 < p^2 < m_b^2 - 2 m_b \Lambda_{QCD}$. 
Beyond this limit the light-cone expansion breaks down,
while the kinematically allowed  momentum range extends up to
$p_{max}^2=(m_B-m_\pi)^2$.
For predictions at large momentum transfer, one may rely on
phenomenological assumptions or models like vector meson dominance.
According to this particular model the form factor $f^+$ should exhibit 
a single pole behaviour:
\bq
f^+(p^2) & = & \frac{f_{B^\ast} g_{B^\ast B \pi}}
{2m_{B^\ast} ( 1-p^2/m_{B^\ast}^2)},
\eq
$m_{B^*}$, $f_{B^*}$, and $g_{B^*B\pi}$ being the mass, decay constant
and coupling of the vector ground state, respectively. 
It is worth noting that the strong coupling $g_{B^\ast B \pi}$ can 
also be calculated from a LCSR based on the same correlation function (3), 
but using a double dispersion relation.
However, for $B \rightarrow \pi$ transitions the single pole model 
cannot be expected to provide a sufficiently accurate and complete 
description \cite{K2000}. Although several improvements have been implemented, 
a solution which is free of additional assumptions has still been missing.
In section I, we discuss a new approach \cite{WY} which is designed to cover 
the whole kinematical range of momentum transfer. 

In the past years, the LCSR have further been refined by including QCD 
corrections to the leading twist 2 contributions
\cite{KRWY2,BBB,B,Khodjamirian:1999hb,Fzero}.
As the most recent example \cite{B,Fzero}, 
the QCD corrections to the scalar form factor 
$f^0$ will be presented in section II.

We note in passing that T. Huang, Z.H. Li and X.Y. Wu \cite{chiral}
have suggested to employ a chiral correlation function, 
in order to reduce the numerical impact of higher 
twist contributions.
Very recently, A. Khodjamirian \cite{Bpipi} extended the LCSR technique 
to the nonleptonic decay $B \rightarrow \pi \pi$.
Finally, also the various updates of the pion wave 
functions \cite{Schm,Bakulev:2001pa,Bakulev:2001ud} should be mentioned
here.

\section{New method of calculating \lowercase{$f^+$}}
In this section we review a new method suggested in \cite{WY} 
for calculating heavy-to-light form factors.
The method is an extension of LCSR and 
is based on first principles. 
The main idea is to use a combination of double and single dispersion relations. 
We rewrite the usual correlation function as
\bq
F_\mu(p,q) &=& F(p^2,(p+q)^2) q_\mu + \tilde{F}(p^2,(p+q)^2) p_\mu, 
\eq
and focus on the invariant amplitude
$F(p^2,(p+q)^2)$. In the following,  we use the definitions 
\bq
\sigma(p^2,s_2)  =  \frac{1}{\pi} \; \mbox{Im}_{s_2} 
\; F(p^2,s_2), \quad 
\rho(s_1,s_2) =  \frac{1}{\pi^2} \; \mbox{Im}_{s_1} \; 
\mbox{Im}_{s_2} \; F(s_1,s_2). 
\eq
The standard sum rule for the form factor $f^+(p^2)$ is obtained 
 by writing a single dispersion relation 
for $F(p^2,(p+q)^2)$ in the $(p+q)^2$-channel, inserting the
hadronic representation for $\sigma(p^2,s_2)$ and Borelizing in $(p+q)^2$:
\bq
\label{eqa1}
{\cal B}_{(p+q)^2} F & = & {\cal B}_{(p+q)^2} 
\left( \frac{2 m_B^2 f_B f^+(p^2)}{m_B^2-(p+q)^2} + 
\int\limits_{s_2>s_0} ds_2 \frac{\sigma^{hadr}(p^2,s_2)}{s_2-(p+q)^2} 
\right).
\eq
Note that any subtraction terms which 
might appear vanish after Borelization.
 Similarly, the standard light-cone sum rule for the coupling 
$g_{B^\ast B \pi}$ is obtained from a double dispersion
relation:
\bq
\label{eqa2}
{\cal B}_{p^2} {\cal B}_{(p+q)^2} F & = &  
{\cal B}_{p^2} {\cal B}_{(p+q)^2} \left( \frac{m_B^2 m_{B^\ast} f_B f_{B^\ast} g_{B^\ast B \pi}}
{(p^2 - m_{B^\ast}^2)( (p+q)^2 - m_B^2)} \right. \nonumber \\
& & \left. + \int\limits_{\Sigma} ds_1 ds_2 \frac{\rho^{hadr}(s_1,s_2)}{(s_1-p^2)(s_2-(p+q)^2)} \right),
\eq
where $\Sigma$ denotes the integration region 
defined by $s_1>s_0$, $s_2>m_b^2$ and $s_1>m_b^2$, $s_2>s_0$.  

In contrast to the above procedure we suggest to use a dispersion relation 
for $\sigma(p^2,s_2)/(p^2)^l$ in the $p^2$-channel 
(with $l$ being an integer):
\bq
\label{eqa3}
\hspace{-6mm}\sigma(p^2,s_2) & = & - \frac{1}{(l-1)!} \left(p^2\right)^l \frac{d^{l-1}}{ds_1^{l-1}}
\left. \frac{\sigma(s_1,s_2)}{s_1-p^2} \right|_{s_1=0} 
+ \int\limits_{s_1>m_b^2} ds_1 \frac{(p^2)^l}{s_1^l} 
\frac{\rho(s_1,s_2)}{s_1 -p^2},  
\eq
and to replace $\sigma(p^2,s_2)$ in (\ref{eqa1})  
by the r.h.s of (\ref{eqa3})
\footnote{By choosing $l$ large 
enough the dispersion relation (\ref{eqa3}) 
will be convergent.}.
Then, writing a double dispersion 
relation for $F(p^2,(p+q)^2)/(p^2)^l$ and 
comparing it with the previous result, we obtain the sum rule
\bq
\label{eqa6}
f^+(p^2) & = & \frac{1}{2} \frac{(p^2)^l}{(m_{B^\ast}^2)^l} \frac{f_{B^\ast} g_{B^\ast B \pi}}
{m_{B^\ast} \left( 1 -\frac{p^2}{m_{B^\ast}^2} \right)} 
- \frac{1}{(l-1)!} \left( p^2 \right)^l \left. \frac{d^{l-1}}{ds_1^{l-1}} 
\frac{f^+(s_1)}{s_1-p^2} \right|_{s_1=0} \nonumber \\
& & + \frac{1}{2 m_B^2 f_B} \int\limits_{\Sigma'} ds_1 ds_2 
\frac{(p^2)^l}{s_1^l} \frac{\rho(s_1,s_2)}{s_1-p^2} e^{- \frac{s_2-m_B^2}{M^2}},
\eq
where the integration region $\Sigma'$ is defined by $s_1>s_0$ and 
$m_b^2 < s_2 < s_0$. This sum rule is valid in the whole 
kinematical range of $p^2$.  As input we need the 
first $(l-1)$ terms of the Taylor expansion of $f^+(p^2)$ 
around $p^2=0$. These parameters can be obtained 
numerically  from the standard sum rule for $f^+(p^2)$:
\bq
\label{eqa1a}
f^+(p^2) & = & \frac{1}{2m_B^2f_B} \int\limits_{m_b^2}^{s_0} \sigma^{QCD}(p^2,s_
2)
e^{-\frac{s_2-m_B^2}{M^2}}
\eq
following from (\ref{eqa1}). 
We further need the residue at the pole $p^2=m_{B^\ast}^2$, 
which can be obtained from the sum rule (\ref{eqa2}).  
 It should be noted that the parameter $l$ plays a 
similar role as the Borel parameter $M^2$. 
There is a lower limit on $l$ such that the dispersion relation  
 (\ref{eqa3}) converges. 
Going to higher values of $l$ will improve the convergence of the 
dispersion relations
and will suppress higher resonances in the $B^\ast$-channel.
But there is also an upper limit on $l$.  
 The higher the value of $l$, the more derivatives
of $f^+(p^2)$ at $p^2=0$ enter. At some point, one starts probing 
the region $p^2 > m_b^2 - 2 m_b \Lambda_{QCD}$, where the standard sum  
rule (\ref{eqa1a}) breaks down.  
\begin{figure}[tr]
\psfig{figure =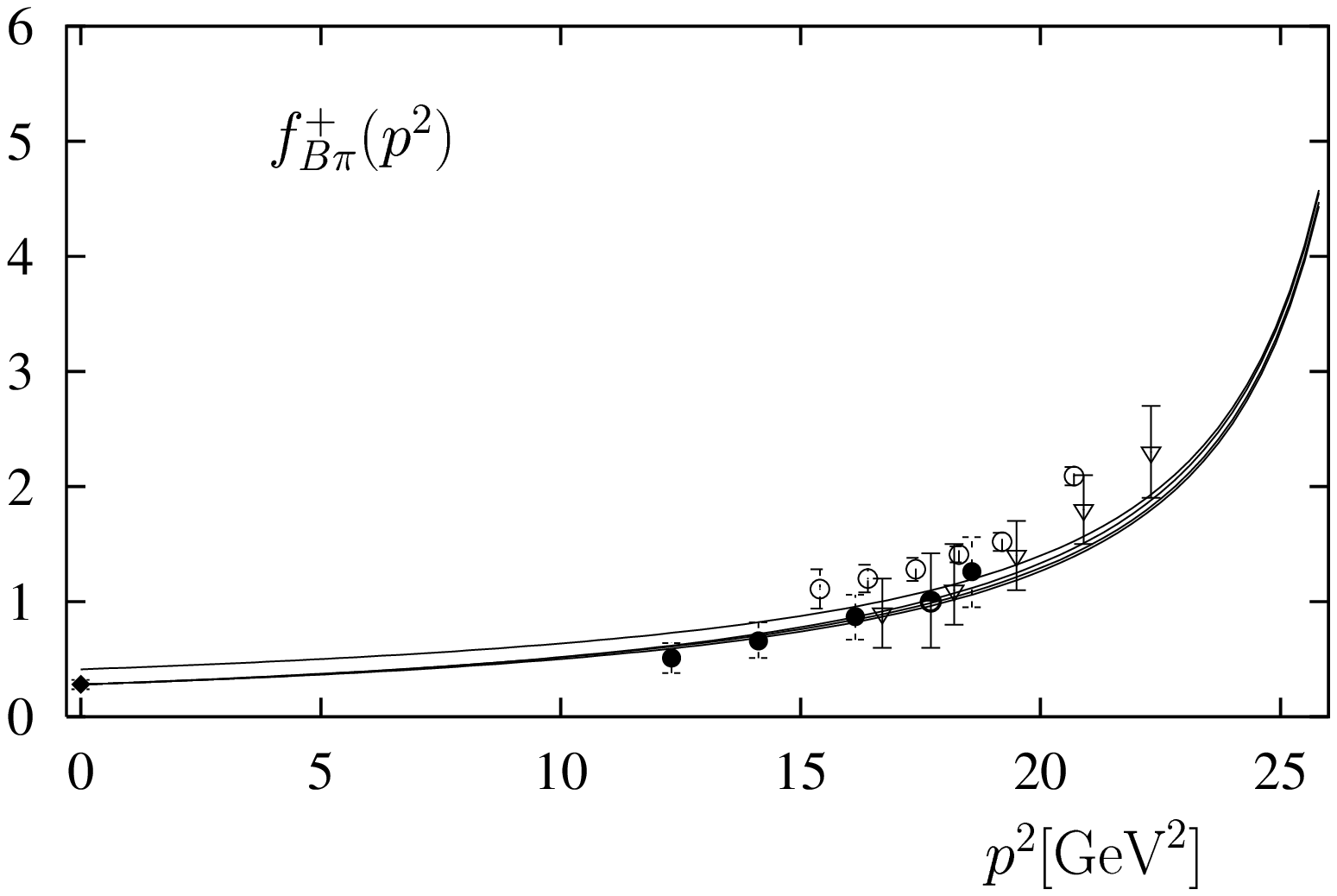,height=1.8in}
\psfig{figure=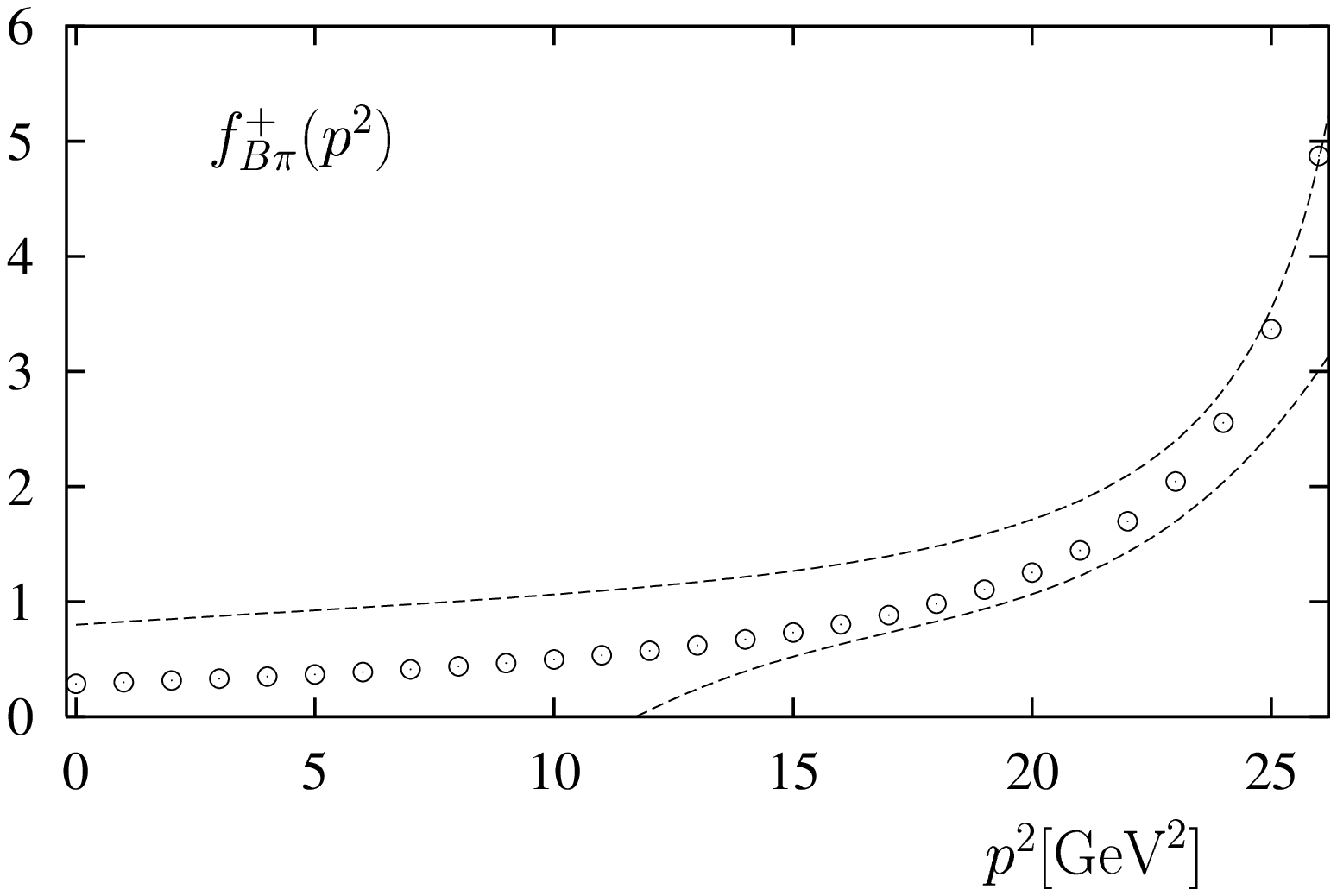,height=1.8in}
\caption{ {\bf Left:} 
LCSR prediction for the $B\to\pi$ 
form factor for $l=0,1,2,3$ in comparison to lattice results
from FNAL (full circles),
UKQCD (triangles), APE (full square), 
JLQCD (open circles), and ELC  (semi-full
circle).  {\bf Right:} 
LCSR prediction on the form factor $f_{B\pi}^+$ (circles)
in comparison to the constraint (dashed) derived by
Boyd and Rothstein.\label{fp1} } 
\end{figure}
Details on the numerical analysis of the new sum rule 
can be found in \cite{WY}. 
Our results are summarized in the convenient parameterization\cite{BK}  
\be\label{paramB}
f_{B\pi}^+(p^2)=\frac{f_{B\pi}^+(0)}{(1-p^2/m_{B^*}^2)(1-
\alpha_{B\pi}p^2/m_{B^*}^2)} ~, 
\ee
with $f^+_{B\pi}(0) = 0.28 \pm 0.05$,
and $\alpha_{B\pi} =  0.4 \pm 0.04~$ in  remarkable agreement with 
$\alpha_{B\pi} =  0.32 \pm^{0.21}_{0.07}$ derived in \cite{K2000}.  
 Fig. \ref{fp1} shows a comparison of (\ref{paramB}) with
recent lattice results \cite{Flynn,lattrev,UKQCD,APE,JLQCD}.
The agreement within uncertainties is very satisfactory.
Finally, the LCSR prediction also obeys the constraints
derived from sum rules for the inclusive
semileptonic decay width in the heavy quark limit \cite{Boyd}.
This is also demonstrated in Fig. \ref{fp1}.

The above results on $f^+_{B \pi}$ can be used 
to calculate the width of the semileptonic decay  
 $B\to \pi \bar{l}\nu_l$ with $l=e,\mu$. 
For the integrated width, one obtains\cite{K2000} 
\be
\Gamma 
= \frac{G^2|V_{ub}|^2}{24\pi^3} \int  dp^2 
(E_\pi^2-m_\pi^2)^{3/2}\left[f^+_{B\pi}(p^2)\right]^2
= (7.3 \pm 2.5) ~|V_{ub}|^2~\mbox{ps}^{-1}~.
\label{elmu}
\ee
Experimentally, combining the branching ratio 
$BR(B^0\to\pi^-l^+\nu_l) = (1.8 \pm 0.6)\cdot 10^{-4}$
with the $B^0$ lifetime
$\tau_{B^0}=1.54 \pm 0.03$ ps 
one gets 
$
\Gamma(B^0\rightarrow \pi ^- l^+ \nu_l) =
(1.17 \pm 0.39)\cdot10^{-4}~\mbox{ps}^{-1}~.
$
From that and (\ref{elmu})
one can then determine the quark mixing parameter $|V_{ub}|$.
The result is
\be
|V_{ub}|=(4.0 \pm 0.7 \pm 0.7)\cdot 10^{-3}
\label{result}
\ee
with the experimental error 
and theoretical uncertainty given in this order.
Using the result analogous to (\ref{paramB})
for the $D\to\pi$ transition one obtains \cite{K2000} 
$\alpha_{D\pi} = 0.01^{+0.11}_{-0.07}~$  and  
$f^+_{D\pi}(0) = 0.65 \pm 0.11,$   
which nicely agrees with lattice estimates,  
 for example, the world average \cite{lattrev} 
$f^+_{D\pi}(0)= 0.65 \pm 0.10 ~,$
or the most recent APE result \cite{APE}, 
$f^+_{D\pi}(0)= 0.64 \pm 0.05 ^{+.00}_{-.07}$.
For more details one should consult\cite{K2000}.

\section{The scalar form factor \lowercase{$f^0$}}
The form factor $f^{0}$ is usually defined through the matrix element
\begin{eqnarray}\label{F0}
p^{\mu}\langle\pi (q)|\bar{u}\gamma_\mu b |B(p+q)\rangle 
= f^0(p^2)(m_B^2-m_{\pi}^2),
\end{eqnarray}
and related to the form factors $f^+$ and $f^-$ as shown in (\ref{scalarff}).
In order to determine $f^0$ from sum rules
it is advantageous to consider  
$f^+$ and $f^+ +f^-$. The sum rule for $f^+$ has been discussed in the 
previous section, the sum rule for $f^+ +f^-$ is schematically given by 
\bq
\label{SumRule}
f^+(p^2) + f^-(p^2) & = & -\frac{m_bf_\pi}{\pi \, m_B^2\,f_B}
\int\limits_{m_b^2}^{s_0}\!\!ds \int\limits_0^1\!\! du 
\,\exp\left(-\frac{s-m_B^2}{M^2}\right)
\nonumber\\
& &\varphi_\pi (u)\,
\mbox{Im}\,\tilde T_{QCD}(p^2,s,u,\mu).
\eq
In the above, only the leading twist 2 contribution is shown,
$\tilde T_{QCD}$ being the corresponding hard scattering
amplitude, $\varphi_\pi (u)$ the pion distribution amplitude,
and $M$ the Borel mass parameter.   
The complete expressions of $\tilde T_{QCD}$ in LO and NLO can be found in 
\cite{KRW} and \cite{Fzero}, respectively. 
Below, we quote
the leading twist-2 QCD correction to the imaginary  
part of the hard amplitude \cite{Fzero}: 
\begin{figure}[tr]
\psfig{figure = 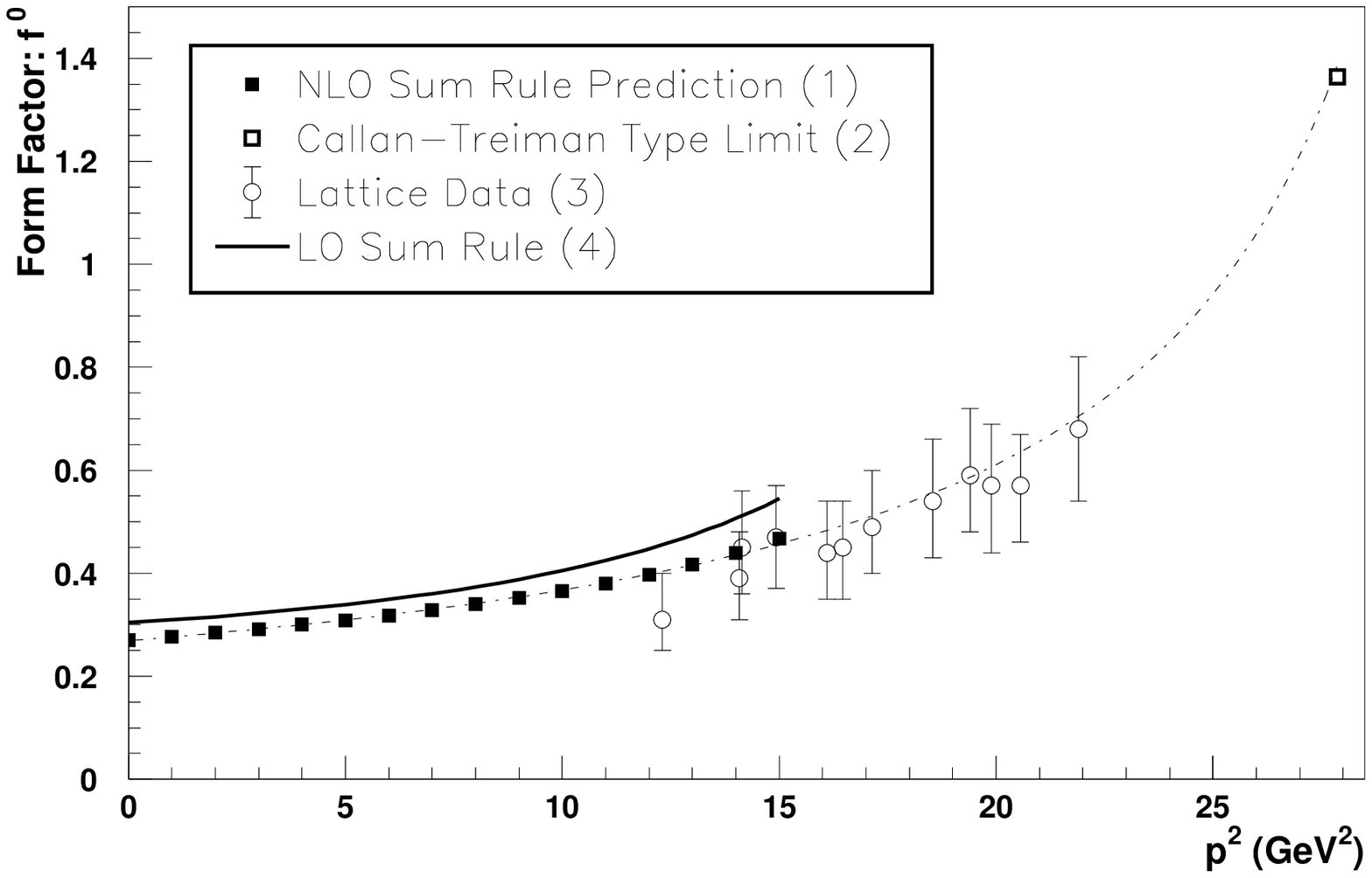,height=2.3in}
\hspace{5mm}
\psfig{figure = 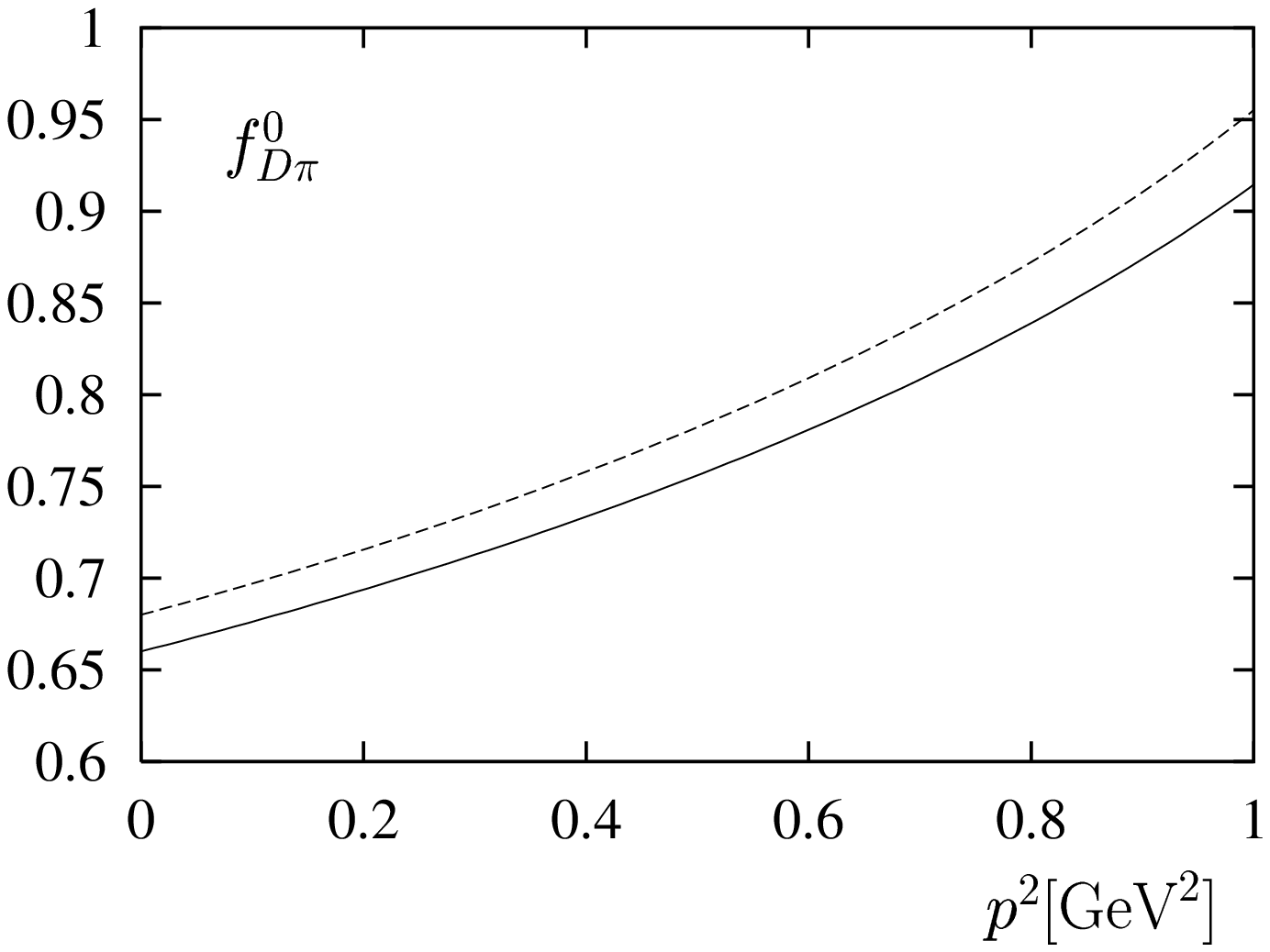,height=1.8in}
\caption{ {\bf Left:}  
NLO LCSR prediction for $f_{B\pi}^0$ (solid squares) 
and extrapolation to the PCAC constraint shown 
by the empty square (dashed line). Also shown are the 
LO LCSR result (solid line) and the UKQCD lattice data (empty circles).
{\bf Right:} LCSR predictions on the form factor 
$f^0_{D\pi}$ in NLO (solid line) and  LO (dashed line). \label{fzero}}
\end{figure}
\begin{eqnarray}\label{ImaginaryPart}
&&\frac{1}{\pi}\mbox{Im}\tilde T_{QCD}(s_1,s_2,u,\mu) =
\Big(\frac{C_F\alpha_s(\mu)}{2\pi}\Big)
\Theta(s_2 - m_b^2) \frac{m_b}{s_2 - s_1}
\nonumber\\ 
&&\Bigg\{ \Theta(u - u_0)
\Bigg[-\frac{(1-u)(u-u_0)(s_2-s_1)^2}{2 u \rho^2} - \frac{1}{u(1-u)}
\left(\frac{m^2}{\rho}-1\right)\Bigg]\\ \nonumber
&+&\delta(u-u_0) \frac{1}{2 u} \Bigg[\frac{(s_1-m_b^2)^2}{s_1^2}
\ln\left(1-\frac{s_1}{m_b^2} \right)
+\frac{m_b^2}{s_1} - 1 \Bigg] 
-\frac{1}{1-u}\left(1-\frac{m_b^2}{s_2}\right)
\Bigg\}
\end{eqnarray}
with $u_0 = \frac{m_b^2 - s_1}{s_2 - s_1}$.
In Fig. \ref{fzero} (left), the resulting form factor $f^0_{B\pi}$ is 
plotted together 
with the UKQCD lattice results \cite{Flynn}. 
It is interesting to see that the radiative effects improve the 
agreement between
the  lattice and the LCSR calculations.  
Also shown in Fig. \ref{fzero} (right) is the LCSR prediction for the 
form factor $f^0_{D\pi}$. 

To conclude, we have discussed 
two improvements of the QCD light-cone sum rules for 
exclusive $B$ annd $D$ decay amplitudes:
firstly, a way to get LCSR predictions for heavy-to-light 
form factors in the complete kinematical
range of momentum transfer without relying on phenomenological models 
such as the single pole model,   
secondly, the inclusion of NLO effects in the LCSR for the 
scalar form factor $f^0$.

{\bf Acknowledgments:} 
O.Y. acknowledges support from the US Department of Energy.
R.R. acknowledges support from the 
Bundesministerium f\"ur Bildung und Forschung (BMBF), Bonn,
Germany, under contract number 05HT9WWA9.


\end{document}